\documentclass[12pt,reqno]{amsart}
\usepackage{graphicx}
\graphicspath{{./art/}}
\usepackage[letterpaper,margin=1in]{geometry}
\usepackage{indentfirst,csquotes,setspace}

\usepackage{amssymb,amsthm,amsmath,mathrsfs}
\usepackage{xcolor,paralist,titlesec,fancyhdr,etoolbox}

\usepackage[ruled,linesnumbered]{algorithm2e}
\SetAlgoNlRelativeSize{0}
\usepackage{subfigure}
\usepackage{enumerate}
\usepackage{booktabs}
\usepackage{threeparttable}
\usepackage{url}
\usepackage{multirow}
\usepackage[authoryear,round]{natbib}
\setcitestyle{authoryear,round,aysep={,},citesep={;}}
\usepackage{float}
\usepackage{placeins}
\usepackage[hypertexnames=false]{hyperref}

\theoremstyle{plain}
\newtheorem{theorem}{Theorem}
\newtheorem{lemma}{Lemma}
\newtheorem{proposition}{Proposition}

\newtheorem{assumption}{Assumption}
\theoremstyle{definition}

\newtheorem{example}{Example}
\theoremstyle{remark}
\newtheorem{remark}{Remark}

\titleformat{\section}[hang]
  {\normalfont\Large\bfseries}{\thesection}{0.8em}{}
\titlespacing*{\section}{0pt}{3.5ex plus 1ex minus .2ex}{2ex plus .2ex}
\titleformat{\subsection}[hang]
  {\normalfont\bfseries}{\thesubsection}{0.8em}{}
\titlespacing*{\subsection}{0pt}{2.5ex plus .8ex minus .2ex}{1ex plus .2ex}
\titleformat{\subsubsection}[hang]
  {\normalfont\itshape}{\thesubsubsection}{0.8em}{}
\titlespacing*{\subsubsection}{0pt}{2ex plus .5ex minus .2ex}{.8ex plus .2ex}
\makeatletter
\renewcommand{\sectionmark}[1]{%
  \begingroup
  \edef\x{\endgroup
    \noexpand\markboth
      {\number\c@section\quad\noexpand\MakeUppercase{\unexpanded{#1}}}%
      {\number\c@section\quad\noexpand\MakeUppercase{\unexpanded{#1}}}}%
  \x}
\providecommand{\@secnumpunct}{.\quad}
\makeatother

\makeatletter
\patchcmd{\@settitle}{\uppercasenonmath\@title}{}{}{}
\patchcmd{\@setauthors}{\MakeUppercase{\authors}}{\authors}{}{}
\makeatother

\hypersetup{hidelinks}

\begin{document}
\setstretch{1.8}
\raggedbottom
\title{Fréchet Oblique Forest for Non-Euclidean Regression}
\author[J. Chen, R. Qiu, and Z. Yu]{%
Jiaye Chen, Rui Qiu, and Zhou Yu\\
\textnormal{\footnotesize School of Statistics, KLATASDS-MOE, East China Normal University}\\
\textnormal{\footnotesize Shanghai 200062, China}}
\date{}

\begin{abstract}
We propose Fréchet oblique forests (FOF) for non-Euclidean regression with predictors and responses in metric spaces. An Alexandrov inner-product score defines directional splits in metric predictor spaces and recovers ordinary oblique splits for Euclidean predictors. Candidate splits are evaluated through pairwise response distances, avoiding repeated Fréchet mean computations during split search. Forest predictions minimize a weighted Fréchet criterion obtained by averaging leaf weights across trees. For compatible tree rules, we establish integrated consistency and polynomial risk bounds under explicit approximation and entropy conditions. Simulations with matrices, distributions, spherical data and mixed predictors, together with an application to taxi-flow networks, demonstrate competitive predictive performance.
\end{abstract}


\maketitle

\bigskip

\section{Introduction}

Regression increasingly involves objects such as probability distributions,
covariance matrices and networks as both predictors and responses. Network
forecasting, for example, may use past networks and periodic covariates
as predictors with different geometries. These choices require particular care in metric
spaces where a linear structure or a natural ordering may not be valid.
For metric-space-valued $X\in\mathcal M_X$ and $Y\in\mathcal M_Y$, the
conditional Fr\'echet mean provides the target
\begin{equation}
	f^\star(x)\in\operatorname*{argmin}_{\omega\in\mathcal M_Y}
	M(x,\omega),\qquad
	M(x,\omega)=\mathbb E\{d_Y^2(Y,\omega)\mid X=x\}.
	\label{eq:conditional-frechet-mean}
\end{equation}
When $\mathcal M_Y=\mathbb R^d$ with the Euclidean metric, this target is
the conditional mean.

For Euclidean predictors, \citet{peterson2019frechet} developed global
and local Fr\'echet regression for object responses. Random forests
\citep{Breiman2001} define adaptive neighbourhoods through leaf weights
\citep{AtheyTibshiraniWager2019}. \citet{qiu2024random} use these weights
for local Fr\'echet regression. Comparatively few methods also allow
object-valued predictors. \citet{bhattacharjee2025nonlinear} provide a
global nonlinear approach using kernel embeddings and weak conditional
expectation. Extending forests to this setting also requires alternatives
to coordinatewise thresholds: object predictors need not have natural
scalar coordinates or an ordering. \citet{capitaine2024frechet} use
Voronoi partitions of individual object-valued predictors. Each split
compares distances to two representatives obtained by clustering or random
sampling. Split selection is based on the reduction in response
Fr\'echet variance.

In flat coordinates, two-representative Voronoi rules can already define
oblique boundaries. The fitted partition depends on how representative
pairs are generated and which predictor component is used at each split.
When representatives are obtained by clustering, the candidate partitions
reflect the predictor distribution. They need not align with the
directions along which the response changes. A response may also depend
jointly on several object components. A joint split can represent a boundary defined by summed directional
scores, whereas componentwise splitting may require several cuts to
approximate it. Splits along predictive directions within and across
objects may thus capture the same relationship with fewer leaves.

Oblique trees in Euclidean regression address a similar problem.
Relationships driven by a few linear combinations of covariates can
require many coordinatewise cuts. Oblique splits can follow those
combinations directly \citep{BreimanFriedmanOlshenStone1984}.
Recent theory relates prediction risk to ridge-function approximation
\citep{CattaneoChandakKlusowski2024}. For oblique Mondrian forests,
risk bounds quantify the role of alignment with relevant predictor
subspaces \citep{OReilly2026}. \citet{ZhanLiuXia2026} establish
consistency for oblique trees and their random forests. A general metric
space, however, need not carry an inner product, so the Euclidean score
$\beta^\top x$ is not directly available. Oblique splitting in this setting
requires a geometric substitute and a criterion for object responses.

We construct Fr\'echet oblique trees and forests using the Alexandrov
inner-product score employed by \citet{lin2025binary}.
On geodesic predictor spaces where this score is defined, it provides a
geometric analogue of $\beta^\top x$. Pairwise response dispersion guides
split selection without repeated Fr\'echet mean computations, and weighted
Fr\'echet means provide predictions. Our contributions are threefold.
First, the resulting framework supports object-on-object regression without
requiring extrinsic embeddings. For Euclidean predictors and responses,
it recovers affine oblique splits and ordinary forest averaging,
respectively. Second, the product-space construction combines component
scores in a single split, allowing objects with different geometries to
contribute jointly to a predictive boundary. Third, for tree rules
satisfying a split-compatibility condition, we establish integrated
consistency under suitable regularity assumptions. Under stronger
assumptions, we derive polynomial bounds on metric prediction risk.

\section{Methodology}
\label{sec:methodology}
\subsection{Oblique trees in Euclidean spaces}

At a node $t\subseteq\mathbb R^p$, an oblique split sends $x$ left if
$\beta^\top x\le c$ and right otherwise. The direction $\beta$ determines
the orientation, and $c$ determines the position. Write
$\mathcal I_t=\{i:X_i\in t\}$ and $n_t=|\mathcal I_t|$ for the node's
observation indices and sample size. For scalar responses, let
$\bar Y_t=n_t^{-1}\sum_{i\in\mathcal I_t}Y_i$ and define the empirical least-squares variance
\[
 \widehat V_t^{\mathrm{LS}}=\frac1{n_t}
 \sum_{i\in\mathcal I_t}(Y_i-\bar Y_t)^2.
\]
A candidate split into children $t_L$ and $t_R$ is evaluated by the
variance reduction
\[
 \Delta_t^{\mathrm{LS}}(\beta,c)
 =\widehat V_t^{\mathrm{LS}}
  -p_{t,L}\widehat V_{t_L}^{\mathrm{LS}}
  -p_{t,R}\widehat V_{t_R}^{\mathrm{LS}},
\]
where $p_{t,L}=n_{t_L}/n_t$ and $p_{t,R}=n_{t_R}/n_t$.
Extending this construction to object data requires a geometric
directional score for the predictors and a dispersion criterion for
the responses.

\subsection{Alexandrov angle and inner product}
\label{sec:alexandrov-angle}

Let $(\mathcal M_X,d_X)$ be a geodesic metric space and fix an anchor $\mu$.
For $x,\beta\ne\mu$, let $\angle_\mu(x,\beta)$ be the Alexandrov angle
between selected minimizing geodesics from $\mu$ to $x$ and $\beta$.
We assume that these angles exist and that the selections make the score
measurable. Define
\begin{equation*}
 h_\mu(x,\beta)=d_X(\mu,x)d_X(\mu,\beta)
                   \cos\angle_\mu(x,\beta),
\end{equation*}
with value zero if either argument equals $\mu$.
The angle is the limit of comparison angles as the two endpoints approach
$\mu$ along the selected geodesics.

On a Riemannian manifold,
\[
 h_\mu(x,\beta)=
 \langle\operatorname{Log}_\mu(x),
        \operatorname{Log}_\mu(\beta)\rangle_\mu,
\]
where the logarithms use the selected minimizing geodesics
\citep{lin2025binary}.
For a product space with squared distance
$\sum_{r=1}^p\lambda_r d_r^2$, $\lambda_r>0$, the score is
$\sum_{r=1}^p\lambda_r h_{\mu_r}(x_r,\beta_r)$ whenever the component
angles exist. The weights $\lambda_r$ set the relative scales of the
component metrics. Joint splitting compares this sum with a single
threshold, so several object components can enter one decision.
Componentwise splitting uses one component score at a time. The component,
direction and threshold are selected using the response-based gain
criterion introduced below.

\subsection{Fr\'echet oblique trees and forests}

Let $(\mathcal M_Y,d_Y)$ be a metric response space and fix an anchor
$\mu\in\mathcal M_X$. A Fr\'echet oblique tree $T$ partitions the
predictor space through a finite sequence of binary splits. At an internal
node $t$, a direction $\beta_t\in\mathcal M_X$ and threshold
$c_t\in\mathbb R$ send $x$ left when $h_\mu(x,\beta_t)\le c_t$
and right otherwise. Write $\mathcal N(T)$ and $\mathcal L(T)$ for the
internal nodes and terminal leaves, respectively, and let $\ell_T(x)$
denote the leaf containing $x$. For Euclidean predictors, these splits
reduce to affine halfspaces.

We observe independent copies $\{(X_i,Y_i)\}_{i=1}^n$ of $(X,Y)$.
Write $P_X$, $P_Y$ and $P_{XY}$ for the marginal and joint laws.
For split selection, we replace the Fr\'echet variance criterion used by
\citet{qiu2024random} with pairwise metric dispersion
\citep{dubey2020functional}. This avoids recomputing the Fr\'echet mean in
each child node for every candidate split. For a response law $P_Y$ with
finite second moment, define $V(P_Y)=\tfrac12\mathbb E\{d_Y^2(Y,Y')\}$, where $Y$ and
$Y'$ are independent draws from $P_Y$. This dispersion need not equal
Fr\'echet variance. Retaining the node notation from Section~2.1, define
\begin{equation*}
 \widehat V_t=\frac1{n_t(n_t-1)}
       \sum_{i<j,\;i,j\in\mathcal I_t}d_Y^2(Y_i,Y_j),\qquad n_t\ge2,
\end{equation*}
and set $\widehat V_t=0$ for $n_t\le1$. The corresponding split gain is
\begin{equation}
 \Delta_t(\beta,c)=\widehat V_t
       -p_{t,L}\widehat V_{t_L}-p_{t,R}\widehat V_{t_R}.
 \label{eq:metric-split-gain}
\end{equation}
% AUTHOR CHECK: document the actual candidate generator, candidate count,
% tuning grids and randomization used for each reported experiment.
% Do not infer these settings from the generic algorithm below.
Algorithm~\ref{alg:optimal_split_node} maximizes this gain over a finite
direction set $\mathcal A_t$ and the corresponding admissible threshold
sets $\mathcal B_t(\beta)$.

\begin{algorithm}[t]
\caption{Nodewise split selection}
\label{alg:optimal_split_node}
\DontPrintSemicolon
\KwIn{Node observations $\mathcal I_t$, anchor $\mu$, candidate directions
$\mathcal A_t$ and minimum child size $n_{\mathrm{leaf}}$.}
\KwOut{A maximizing split and its gain, or a terminal-node decision.}
\ForEach{$\beta\in\mathcal A_t$}{
Compute $s_i=h_\mu(X_i,\beta)$ for $i\in\mathcal I_t$ and sort the
distinct scores as $u_1<\cdots<u_J$.\;
Define $\mathcal B_t(\beta)$ as the set of midpoints
$(u_j+u_{j+1})/2$, $1\le j<J$, that yield at least
$n_{\mathrm{leaf}}$ observations per child.\;
\ForEach{$c\in\mathcal B_t(\beta)$}{
Set $\mathcal I_{t_L}=\{i\in\mathcal I_t:s_i\le c\}$ and
$\mathcal I_{t_R}=\mathcal I_t\setminus\mathcal I_{t_L}$.\;
Compute $\Delta_t(\beta,c)$ from \eqref{eq:metric-split-gain}.\;
}
}
\eIf{$\mathcal B_t(\beta)=\varnothing$ for every $\beta\in\mathcal A_t$}{
Declare the node terminal.\;
}{
Choose
$
 (\widehat\beta_t,\widehat c_t)\in
 \operatorname*{argmax}_{\substack{\beta\in\mathcal A_t, c\in\mathcal B_t(\beta)}}
       \Delta_t(\beta,c).
$
Return $(\widehat\beta_t,\widehat c_t)$ and
$\Delta_t(\widehat\beta_t,\widehat c_t)$.\;
}
\end{algorithm}

For scalar Euclidean responses and $n_t\ge2$, $\widehat V_t$ is the sample variance with
denominator $n_t-1$, whereas $\widehat V_t^{\mathrm{LS}}$ uses $n_t$.
Thus $\Delta_t$ need not select the same split as $\Delta_t^{\mathrm{LS}}$.
Recursion with minimum-size and
minimum-gain stopping rules gives the fitted tree. For a fitted tree $T$
with nonempty leaves, all node sums and counts refer to its training
sample. Its criterion and prediction are
\[
 \widehat M_T(x,\omega)=\frac1{n_{\ell_T(x)}}
       \sum_{i\in\mathcal I_{\ell_T(x)}}d_Y^2(Y_i,\omega),\qquad
 \widehat f_T(x)\in\operatorname*{argmin}_{\omega\in\mathcal M_Y}
       \widehat M_T(x,\omega).
\]
We assume that the prediction criteria attain their minima and use
measurable choices of minimizers.

To form a forest, draw size-$N_n$ subsamples
$\mathcal I^{(1)},\ldots,\mathcal I^{(B_n)}$ uniformly without replacement,
independently of the observations. For the $b$th fitted tree $T^{(b)}$, write
$\ell_b(x)=\ell_{T^{(b)}}(x)$ and let $n_t^{(b)}$ count its
sampled observations in node $t$. With $\mathbf 1\{A\}$ denoting the
indicator of $A$, define
\begin{equation}
 \alpha_i^{(b)}(x)=
 \frac{\mathbf 1\{i\in\mathcal I^{(b)},\ X_i\in\ell_b(x)\}}
      {n_{\ell_b(x)}^{(b)}},
 \qquad \bar\alpha_i(x)=\frac1{B_n}\sum_{b=1}^{B_n}\alpha_i^{(b)}(x).
 \label{eq:forest-weights-main}
\end{equation}
These weights are nonnegative and sum to one. Writing
$\widehat M^{(b)}=\widehat M_{T^{(b)}}$ for the criterion of
tree $b$, define the forest criterion and prediction by
\begin{equation}
 \begin{aligned}
 \widehat M_n(x,\omega)
 =\frac1{B_n}\sum_{b=1}^{B_n}\widehat M^{(b)}(x,\omega)
  =\sum_{i=1}^n\bar\alpha_i(x)d_Y^2(Y_i,\omega), \qquad
 \widehat f_n(x)\in\operatorname*{argmin}_{\omega\in\mathcal M_Y}
                 \widehat M_n(x,\omega).
 \end{aligned}
 \label{eq:forest-weighted-objective}
\end{equation}
The subscript $n$ records the full sample size; dependence on $B_n$ and
other fitting parameters is suppressed. When $B_n=1$, the prediction is
$\widehat f_T$ for $T=T^{(1)}$. With multiple trees, the leaf
weights are averaged before minimization to pool information across trees.
Algorithm~\ref{alg:frechet_oblique_forest} in Appendix~2 gives the full forest procedure.

\section{Theoretical properties}
\label{sec:theory}

We study trees grown to depth $K_n$ on size-$N_n$ subsamples drawn without
replacement, with a fixed deterministic anchor $\mu$. A depth $K_n$ is
admissible when the fitted rule can split every node at depth less than
$K_n$ into two children that both contain observations. This requires $2^{K_n}\le N_n$,
and admissibility also depends on ties and the available splits. The
response space is compact, with diameter $D_Y<\infty$. We use a version
of $M(x,\cdot)$ that is $2D_Y$-Lipschitz for $P_X$-almost every $x$.
Expectations in the risk bounds are over the training sample, subsampling
and split randomization.

\subsection{Consistency of the Fr\'echet oblique forest}
\label{sec:forest-consistency}

Let $S=\operatorname{supp}(P_X)$ denote the support of the predictor
distribution, assumed measurable and of full $P_X$-probability.

For $\omega\in\mathcal M_Y$, put $Z_i(\omega)=d_Y^2(Y_i,\omega)$ and
$M_\omega(x)=M(x,\omega)$. Fix a countable dense subset
$\Omega_0\subseteq\mathcal M_Y$, independently of the data.
The split class is
$\mathcal H_\mu=\{\{x:h_\mu(x,\beta)\le c\}:\beta\in\mathcal M_X,c\in\mathbb R\}$.

\setcounter{assumption}{0}
\renewcommand{\theassumption}{A\arabic{assumption}}
\begin{assumption}[Identification]
\label{ass:identification}
The conditional Fr\'echet minimizer $f^\star(x)$ defined in
\eqref{eq:conditional-frechet-mean} is unique for $P_X$-almost every $x$.
\end{assumption}

\begin{assumption}[Split complexity]
\label{ass:split-complexity}
The score $h_\mu$ is measurable. Identify each set in $\mathcal H_\mu$
with its indicator in $L^1(P_X)$. For the bracketing entropy
$H(\varepsilon)=\log N_{[]}\{\varepsilon,\mathcal H_\mu,L^1(P_X)\}$,
assume at least one of the following bounds holds for all $0<\varepsilon<1$
\begin{equation}
 \textnormal{(i)}\quad H(\varepsilon)\le C_H\log(e/\varepsilon),
 \qquad\textnormal{or}\qquad
 \textnormal{(ii)}\quad H(\varepsilon)\le C_H\varepsilon^{-q},
 \label{eq:A2-bracketing}
\end{equation}
for fixed $C_H<\infty$ and, in (ii), $q>0$.
Each bracket has measurable endpoints $A^-\subseteq A^+$ with
$P_X(A^+\setminus A^-)\le\varepsilon$. The inclusions
$A^-\subseteq A\subseteq A^+$ hold pointwise, or on one common set of
full probability for the entire class.
\end{assumption}

Case (i) applies when the splits can be written as affine halfspaces in
fixed finite-dimensional coordinates on $S$. Examples include Euclidean
predictors, Log--Cholesky coordinates for $d\times d$ symmetric positive
definite (SPD) matrices
\citep{lin2019riemannian}, and fixed-shape Wasserstein laws. Let
$\mathcal W_2(\mathbb R)$ be the space of probability laws on $\mathbb R$
with finite second moment, equipped with the 2-Wasserstein distance $W_2$,
and let $Q_x$ be the quantile function of $x\in\mathcal W_2(\mathbb R)$.
For fixed-shape laws, the quantile functions are $a+bQ_0$, where
$a\in\mathbb R$, $b>0$, and $Q_0$ is a fixed quantile function.
These three representations use $p$, $d(d+1)/2$, and at most two
coordinates, respectively. For a connected complete $p$-dimensional
Riemannian predictor space, the Borel minimizing-log selection
$Z_\mu=\operatorname{Log}^{\mathrm{sel}}_\mu$ gives coordinates in
$T_\mu\mathcal M_X$, with
\[
 h_\mu(x,\beta)=\langle Z_\mu(x),Z_\mu(\beta)\rangle_\mu.
\]
Here $\langle\cdot,\cdot\rangle_\mu$ is the Riemannian inner product.
This includes finite-dimensional hyperbolic and spherical predictors
under the selection described in Lemma~\ref{lem:ridge-density}.

Write $D$ for the coordinate dimension and $\mathcal H_D^{\mathrm{aff}}$
for the class of sets $\{z:a^\top z\le b\}$, with $a\in\mathbb R^D$ and
$b\in\mathbb R$. For a suitable
measurable coordinate map $T:S\to\mathbb R^D$, the splits restricted to $S$
are pullbacks of affine halfspaces. The bracketing bound of
\citet[Corollary~6]{DuttaGhoshMoran2022} gives
\[
 \{A\cap S:A\in\mathcal H_\mu\}
 \subseteq \{T^{-1}(A):A\in\mathcal H_D^{\mathrm{aff}}\},
 \qquad
 H(\varepsilon)\le C(D+1)\log(C/\varepsilon)
 \le C_H\log(e/\varepsilon).
\]
\citet[Corollary~3.8]{HaghtalabRoughgardenShetty2020} use logarithmic
bracketing bounds for polytopes and polynomial threshold classes in
smoothed online learning.

In case (ii), for $\mathcal M_X=\mathcal W_2(\mathbb R)$, the
nested-subspace condition \eqref{eq:wass-hpfd} of
Proposition~\ref{prop:wass-hpfd} gives
\[
 H(\varepsilon)\le C\varepsilon^{-1/r}\log(C/\varepsilon)
 \le C_q\varepsilon^{-q},\qquad q>1/r.
\]
The condition requires $\Pr(Q_X\notin V_m)\le C_0m^{-r}$ for nested
subspaces $V_m$ of dimension at most $m$. If $X\in\mathcal P_2([a,b])$ almost surely for a fixed bounded
interval $[a,b]$, where $\mathcal P_2([a,b])$ denotes the probability laws
on $[a,b]$, then $S\subseteq\mathcal P_2([a,b])$ is compact in $W_2$. This supplies the compactness condition in
Lemma~\ref{lem:ridge-density}.

\citet[Assumption~A(E1)]{Han2021SetStructured} uses a polynomial
$L^1(P)$ bracketing bound of the form in (ii) for empirical processes
indexed by measurable sets. In metric-space classification,
\citet[Assumption~1]{lin2025binary} use analogous logarithmic and
polynomial bounds for the covering entropy of predictor-space balls.

\begin{remark}[Random finite quantile splines]
\label{rem:random-quantile-splines}
For $r>0$, let
$\Pr(M_Q=2^\ell)=(1-2^{-r})2^{-r\ell}$, $\ell\ge0$, and, conditional on
$M_Q$, draw $(W_i)_{i=1}^{M_Q}\sim\operatorname{Dirichlet}(1,\ldots,1)$.
Linearly interpolate the knots
$Q_X(j/M_Q)=\sum_{i=1}^jW_i$, $0\le j\le M_Q$.
Let $V^{(\ell)}$ be the piecewise-linear spline space on
$\{j/2^\ell:0\le j\le2^\ell\}$, and set
$V_m=V^{(\lfloor\log_2(m-1)\rfloor)}$ for $m\ge2$, with $V_1=\{0\}$.
This family satisfies \eqref{eq:wass-hpfd}, with $\dim V_m\le m$ and
$\Pr(Q_X\notin V_m)\le C_rm^{-r}$. The law has compact predictor support
in $\mathcal P_2([0,1])$ and satisfies $\mathbb E Q_X(t)=t$, with fixed
anchor $\mu=\operatorname{Unif}[0,1]$.
\end{remark}

Consider a tree $T_{m,K}$ fitted to $m$ independent observations.
Retain $\mathcal I_t$, $n_t$, $p_{t,L}$ and $p_{t,R}$ for its nodes,
and put $w_t=n_t/m$. For the node average
$\bar Z_t(\omega)=n_t^{-1}\sum_{i\in\mathcal I_t}Z_i(\omega)$,
the least-squares gain for $Z_i(\omega)$, weighted by $w_t$, is
\begin{equation*}
 \Gamma_{t,\omega}(\beta,c)
 =w_tp_{t,L}p_{t,R}\{\bar Z_{t_L}(\omega)-\bar Z_{t_R}(\omega)\}^2.
\end{equation*}
Let $\mathcal S_t$ consist of the pairs $(\beta,c)\in\mathcal M_X\times
\mathbb R$ for which the split $\{x:h_\mu(x,\beta)\le c\}$ assigns at least
one observation of node $t$ to each child. Let $T_{m,k}$ be the
depth-$k$ truncation of $T_{m,K}$, with the root at depth zero, and let
$\mathscr F_k$ be the sigma-field generated by the training data and the
random choices that determine $T_{m,k}$.

\begin{assumption}[Conditional split compatibility]
\label{ass:split-compatibility}
For each $\omega\in\Omega_0$, a deterministic $c_\omega>0$, independent of
$m,k,t$, satisfies that almost surely
\begin{equation*}
 \mathbb E\{\Gamma_{t,\omega}(\widehat\beta_t,\widehat c_t)
                  \mid\mathscr F_k\}
 \ge c_\omega\sup_{(\beta,c)\in\mathcal S_t}\Gamma_{t,\omega}(\beta,c) 
 \qquad t\in\mathcal L(T_{m,k}),\quad k<K.
\end{equation*}
\end{assumption}

This assumption adapts the approximate split-optimization framework of
\citet[Section~2.2 and Assumption~1]{CattaneoChandakKlusowski2024}
to conditional expected gains for the scalar responses $Z_i(\omega)$.
For the common tree rule, a sufficient condition for each fixed $\omega$ is
\[
 \Pr\!\left\{\Gamma_{t,\omega}(\widehat\beta_t,\widehat c_t)
 \ge\kappa_\omega\sup_{(\beta,c)\in\mathcal S_t}
       \Gamma_{t,\omega}(\beta,c)\mid\mathscr F_k\right\}
 \ge\pi_\omega>0
 \quad\Longrightarrow\quad c_\omega=\kappa_\omega\pi_\omega,
\]
where $\kappa_\omega>0$ and $\pi_\omega$ are deterministic and independent
of sample size, depth and node. Exact maximization of the scalar gain gives
$\kappa_\omega=\pi_\omega=1$.

Write $I_\beta$ for the smallest interval containing the range of
$h_\mu(\cdot,\beta)$ on $S$.
For a real function $u$ on an interval $I$, $\mathrm{TV}_I(u)$ denotes
its total variation, namely the supremum of
$\sum_{r=1}^{R-1}|u(v_{r+1})-u(v_r)|$ over all finite increasing sequences
$v_1<\cdots<v_R$ in $I$. We use the ridge library
\begin{equation*}
 \mathcal G_\mu=\left\{g=c+\sum_{j=1}^J
    g_j\{h_\mu(\cdot,\beta_j)\}:J<\infty,
    \ \sum_{j=1}^J\mathrm{TV}_{I_{\beta_j}}(g_j)<\infty\right\}.
\end{equation*}
Here $c\in\mathbb R$, $J$ is a finite positive integer,
$\beta_j\in\mathcal M_X$, and each $g_j$ is a real function of bounded
variation on $I_{\beta_j}$.

\begin{theorem}[Forest consistency]
\label{thm:fof-consistency}
Suppose Assumptions~\ref{ass:identification}--\ref{ass:split-compatibility}
hold and the ridge-density hypotheses of Lemma~\ref{lem:ridge-density}
are satisfied. Let $N_n\to\infty$ and let admissible $K_n\to\infty$ satisfy,
in cases (i) and (ii), respectively,
\begin{align*}
 &2^{K_n}(C_H+1)\log(eN_n)/N_n\longrightarrow0,\\
 &2^{K_n}N_n^{-1/(q+1)}\longrightarrow0.
\end{align*}
For any deterministic integer sequence $B_n\ge1$,
\begin{equation*}
 \mathbb E\!\left[\int d_Y^2\{\widehat f_n(x),f^\star(x)\}
                         \,dP_X(x)\right]\longrightarrow0.
\end{equation*}
\end{theorem}

Lemma~\ref{lem:ridge-density} covers compact predictor supports in
one-dimensional Wasserstein space, Log--Cholesky SPD space and
finite-dimensional complete Riemannian manifolds with a fixed minimizing-log
selection.

\subsection{Polynomial convergence rate}
\label{sec:forest-rate}

The polynomial rate uses the following conditions in addition to
Assumption~\ref{ass:split-complexity}.

For a real function $g$ on the predictor space, write
$\|g\|_{P,2}^2=\int g(x)^2\,dP_X(x)$. For a random variable $U$, let
$\|U\|_{L^r}=(\mathbb E|U|^r)^{1/r}$, $r\ge1$.

For a node $t$, let $I_\beta(t)$ be the smallest interval containing
the range of $h_\mu(\cdot,\beta)$ on $t\cap S$.
The global variation $\mathcal V_\mu(g)$ is the infimum of the
total-variation sum over all representations of $g$. Within a node, define
\begin{equation*}
 \mathcal V_\mu(g;t)=
 \inf_{g=c+\sum_jg_j\{h_\mu(\cdot,\beta_j)\}}
       \sum_j\mathrm{TV}_{I_{\beta_j}(t)}(g_j).
\end{equation*}
Constants have zero variation.

\begin{assumption}[Squared response distance of negative type]
\label{ass:negative-type}
For every $r\ge1$, $y_1,\ldots,y_r\in\mathcal M_Y$ and
$b_1,\ldots,b_r\in\mathbb R$ with $\sum_i b_i=0$,
\begin{equation*}
 \sum_{i,l=1}^r b_i b_l d_Y^2(y_i,y_l)\le0.
\end{equation*}
Fix $o\in\mathcal M_Y$. The kernel
$k_o(y,\omega)=\{d_Y^2(y,o)+d_Y^2(\omega,o)-d_Y^2(y,\omega)\}/2$
has a Hilbert representation $k_o(y,\omega)=
\langle\Phi(y),\Phi(\omega)\rangle_{\mathbb H}$ with $\Phi(o)=0$ and
$d_Y(y,\omega)=\|\Phi(y)-\Phi(\omega)\|_{\mathbb H}$
\citep{SejdinovicSriperumbudurGrettonFukumizu2013}.
Take $\mathbb H=\overline{\operatorname{span}}\Phi(\mathcal M_Y)$.
The image of the compact response space is compact, so this Hilbert space
is separable. Fix a deterministic orthonormal basis $(e_j)_{j\in\mathcal J}$,
where $\mathcal J$ is finite or countably infinite, and put
\begin{equation*}
 \phi_j(y)=\langle\Phi(y),e_j\rangle_{\mathbb H},\qquad
 a_j=\sup_{y\in\mathcal M_Y}|\phi_j(y)|>0,\qquad
 z_j(y)=\phi_j(y)/a_j.
\end{equation*}
 Thus $|z_j|\le1$; the embedding and basis are fixed independently of the data.
\end{assumption}
Assumption~\ref{ass:negative-type} holds for compact subsets
of SPD matrices with the Log--Cholesky metric \citep{lin2019riemannian},
of spheres with chordal distance as subsets of Euclidean space
\citep{SejdinovicSriperumbudurGrettonFukumizu2013}, and of
$\mathcal W_2(\mathbb R)$ with $W_2$ through its $L^2$ quantile embedding
\citep{PanaretosZemel2019}.

\begin{assumption}[Coordinatewise split compatibility]
\label{ass:coordinate-compatibility}
For $Z_{ij}=z_j(Y_i)$ and
$\bar Z_{t,j}=n_t^{-1}\sum_{i\in\mathcal I_t}Z_{ij}$, define
\[
 \Gamma_{t,j}(\beta,c)=w_tp_{t,L}p_{t,R}
 \{\bar Z_{t_L,j}-\bar Z_{t_R,j}\}^2.
\]
For each $j\in\mathcal J$, there is a deterministic $c_j\in(0,1]$, independent of
sample size, depth and node, such that the same tree rule satisfies
\begin{equation*}
 \mathbb E\{\Gamma_{t,j}(\widehat\beta_t,\widehat c_t)
             \mid\mathscr F_k\}
 \ge c_j\sup_{(\beta,c)\in\mathcal S_t}\Gamma_{t,j}(\beta,c),
 \quad t\in\mathcal L(T_{m,k}),\ k<K,\quad\textnormal{a.s.}
\end{equation*}
\end{assumption}

\begin{assumption}[Coordinatewise local ridge approximation]
\label{ass:coordinate-approximation}
Let $m_j(x)=\mathbb E\{z_j(Y)\mid X=x\}$. There are a common $s>2$ and
deterministic $V_j\ge0$ such that, for each $j\in\mathcal J$, a deterministic sequence
$g_{j,\ell}\in\mathcal G_\mu$, independent of the sample and tree randomization,
satisfies
\begin{equation*}
 \|g_{j,\ell}-m_j\|_{P,2}\xrightarrow[\ell\to\infty]{}0,
 \qquad
 \sup_{\ell\ge1}\mathbb E
 \sum_{t\in\mathcal L(T_{m,k})}
       \mathcal V_\mu(g_{j,\ell};t)^s\le V_j^s
\end{equation*}
for every sample size and $0\le k<K$ under consideration, with the same
approximating sequence throughout.
\end{assumption}
Assumptions~\ref{ass:coordinate-compatibility} and
\ref{ass:coordinate-approximation} adapt the split-gain and local-variation
conditions of \citet{CattaneoChandakKlusowski2024} to response coordinates
while keeping one common tree rule.

\begin{assumption}[Weighted square summability]
\label{ass:coordinate-summability}
For the coordinates and constants above,
\begin{equation*}
 \Lambda=\sum_{j\in\mathcal J}a_j^2\left(1+\frac{V_j^2}{c_j}\right)<\infty.
\end{equation*}
\end{assumption}
Similar square-summability conditions appear in \citet{Krebs2019} for
coordinatewise regularity constants and in \citet{YaoMullerWang2005} for
weighted functional expansions. \citet[Assumptions~1 and~2]{CilibertoRosascoRudi2020}
combine weighted square summability of Hilbert-valued coefficients with a
bound on the effective dimension of the input covariance operator.

\begin{assumption}[Node-size moment bound]
\label{ass:node-size}
For fixed $A<\infty$ and $\nu\ge s/(s-2)$,
\begin{equation*}
 \left\|\max_{t\in\mathcal L(T_{m,k})}w_t\right\|_{L^\nu}
 \le A\,2^{-k},\qquad 0\le k<K.
\end{equation*}
The bound holds for every sample size under consideration; the norm includes
the data and tree randomization.
\end{assumption}

\begin{assumption}[Uniform quadratic growth]
\label{ass:quadratic-growth}
For some fixed $c_0>0$,
\begin{equation*}
 M(x,\omega)-M\{x,f^\star(x)\}
 \ge c_0d_Y^2\{\omega,f^\star(x)\},
 \qquad\omega\in\mathcal M_Y,
\end{equation*}
on a common full-$P_X$ set.
\end{assumption}
For a Hadamard response space, the barycentric variance inequality gives
Assumption~\ref{ass:quadratic-growth} with $c_0=1$
\citep[Proposition~4.4]{Sturm2003}.

\begin{theorem}[Forest polynomial rate]
\label{thm:fof-rate}
Suppose Assumptions~\ref{ass:split-complexity} and
\ref{ass:negative-type}--\ref{ass:quadratic-growth} hold. Let $N_n\to\infty$.
For sufficiently large $n$, suppose the following depths are admissible:
\begin{align*}
 K_n&=\left\lfloor\frac{s}{s+2}
       \log_2\{N_n/\log(eN_n)\}\right\rfloor,
       &&\textnormal{in case (i)},\\
 K_n&=\left\lfloor\frac{s}{(q+1)(s+2)}\log_2N_n\right\rfloor,
       &&\textnormal{in case (ii)}.
\end{align*}
For any deterministic integer sequence $B_n\ge1$,
\begin{equation*}
 \mathbb E\!\left[\int d_Y^2\{\widehat f_n(x),f^\star(x)\}
                         \,dP_X(x)\right]
 =\begin{cases}
 O[\{\log(eN_n)/N_n\}^{2/(s+2)}],&\textnormal{in case (i)},\\
 O[N_n^{-2/\{(q+1)(s+2)\}}],&\textnormal{in case (ii)}.
 \end{cases}
\end{equation*}
\end{theorem}

Case (i) has the same order as the risk upper bound of
\citet[Theorem~3.2]{CattaneoChandakKlusowski2024} for fixed-dimensional
Euclidean predictors and bounded scalar responses.
Case (ii) additionally reflects polynomial
split entropy through the factor $q+1$ in the denominator of the exponent.

Both results allow a fixed number of trees. Setting $N_n=n$ and $B_n=1$
recovers the single-tree conclusions.

\section{Numerical experiments}
\subsection{Simulations}

We evaluate FOF in regression problems involving SPD matrices, probability
distributions and spherical data. We present the SPD-to-SPD
experiment here and report the remaining settings in Appendix~3.
The following model, adapted from \citet{lin2025binary} and
\citet{yuan2026frechet}, makes the response depend on the predictor
through the directional score $h_\mu(X,\beta)$.

\begin{example}[SPD predictors and responses]
\label{ex:spd-spd}
Let $\mathcal M_X=\mathcal S_{++}^3$ and
$\mathcal M_Y=\mathcal S_{++}^2$, where $\mathcal S_{++}^d$ denotes the
space of $d\times d$ SPD matrices. Equip both spaces with the
Log--Cholesky metric $d_{LC}$. Fix $\mu=I_3$ and consider two directions,
$\beta=3I_3$ and $\beta=3\Sigma_{\mathrm{AR}}$. Here $I_d$ is the
$d\times d$ identity matrix and
$\Sigma_{\mathrm{AR}}=(0.5^{|j-k|})_{1\le j,k\le3}$.
For each $i=1,\ldots,n$, independently generate
\[
 X_i=\operatorname{Exp}_\mu(S_i),
 \qquad \vec S_i\sim N_6(0,4I_6),
\]
where $\operatorname{Exp}_\mu$ is the Riemannian exponential map for the
Log--Cholesky metric, and $\vec S_i$ gives the coordinates of
$S_i$ in an orthonormal basis of $T_\mu\mathcal M_X$.
Generate the response by
\[
 Y_i=\exp\left\{
 \begin{pmatrix}
  1&0.8h_\mu(X_i,\beta)\\
  0.8h_\mu(X_i,\beta)&1
 \end{pmatrix}
 +0.2 E_i\right\},
\]
where $\exp$ denotes the matrix exponential and $h_\mu$ is the score in
Section~\ref{sec:alexandrov-angle}. The noise matrices $E_i$ are symmetric,
with independent upper-triangular entries
$(E_i)_{11},(E_i)_{22}\sim N(0,1)$ and $(E_i)_{12}\sim N(0,1/4)$.
They are independent across $i$ and independent of the predictors.
\end{example}

We compare FOF with Fr\'echet random forests (FRF)
\citep{capitaine2024frechet} and global nonlinear Fr\'echet regression
(GNLFR) \citep{bhattacharjee2025nonlinear}. For each training sample size
$n=200,500,800$, we use an independent test sample of size $n/4$.
Prediction loss is the squared Log--Cholesky distance.
Table~\ref{tab:spd_spd_mse} reports test mean squared prediction errors
(MSEs), averaged over 100 independent replications.

\begin{table}[htbp]
\centering
\caption{Mean squared prediction errors in Example~\ref{ex:spd-spd}.
Monte Carlo standard deviations are in parentheses; the lowest error
for each $n$ and $\beta$ is in bold}
\label{tab:spd_spd_mse}
\small
\begin{tabular*}{\textwidth}
{@{\extracolsep{\fill}}lcccccc@{}}
\toprule
& \multicolumn{2}{c}{FOF}
& \multicolumn{2}{c}{FRF}
& \multicolumn{2}{c}{GNLFR} \\
\cmidrule(lr){2-3}\cmidrule(lr){4-5}\cmidrule(lr){6-7}
$n$
& $3I_3$
& $3\Sigma_{\mathrm{AR}}$
& $3I_3$
& $3\Sigma_{\mathrm{AR}}$
& $3I_3$
& $3\Sigma_{\mathrm{AR}}$ \\
\midrule
200
& \textbf{1.11} (0.79)
& \textbf{11.58} (19.14)
& 1.79 (1.03)
& 15.43 (22.03)
& 2.17 (1.08)
& 17.77 (21.87) \\
500
& \textbf{0.76} (1.39)
& \textbf{10.53} (22.34)
& 1.39 (1.56)
& 16.05 (25.12)
& 2.35 (1.63)
& 21.27 (26.85) \\
800
& \textbf{0.61} (0.71)
& \textbf{5.36} (6.39)
& 1.23 (0.89)
& 10.32 (9.21)
& 2.29 (1.04)
& 16.71 (11.28) \\
\bottomrule
\end{tabular*}
\end{table}

FOF has the lowest mean MSE across all six combinations of sample size
and direction. For $\beta=3I_3$, it reduces MSE by $38.1$--$50.4\%$
relative to FRF, and both forests outperform GNLFR. Errors are larger
for $\beta=3\Sigma_{\mathrm{AR}}$, with substantially greater variation
across replications for all three methods.

\subsection{Real-data analysis}
\label{sec:taxi}
We analyse 1,416 hourly taxi-flow networks from New York City in
January--February 2019. Each network has a symmetric $3\times3$ weighted
adjacency matrix. Its matrix exponential gives the SPD matrix $S_t$.
Hourly normalization makes these networks describe relative spatial
flow patterns.
For one-hour-ahead prediction, we set
\[
    Y_t=S_t,\qquad
    X_t=
    \bigl(S_{t-1},S_{t-24},S_{t-168},
          C_t^{\mathrm{day}},C_t^{\mathrm{week}}\bigr),
\]
where the SPD predictors represent the preceding hour and the same hour
of the previous day and week. The circular predictors encode time of day
and time of week. We use the Log--Cholesky metric for all SPD objects.

The weekly lag leaves 1,248 usable observations, divided chronologically
into 748 training, 249 validation and 251 test observations.
We select hyperparameters for FOF, FRF and GNLFR by validation MSE,
then refit each method on the combined training and validation sets
(997 observations). Both forests use 100 trees and share the 24-hour
moving-block bootstrap samples, with a sampling fraction of $0.8$.
During testing, we update the lagged
predictors with observed networks and keep the fitted models fixed.

\begin{figure}[hbt!]
\centering
\includegraphics[width=0.85\textwidth]{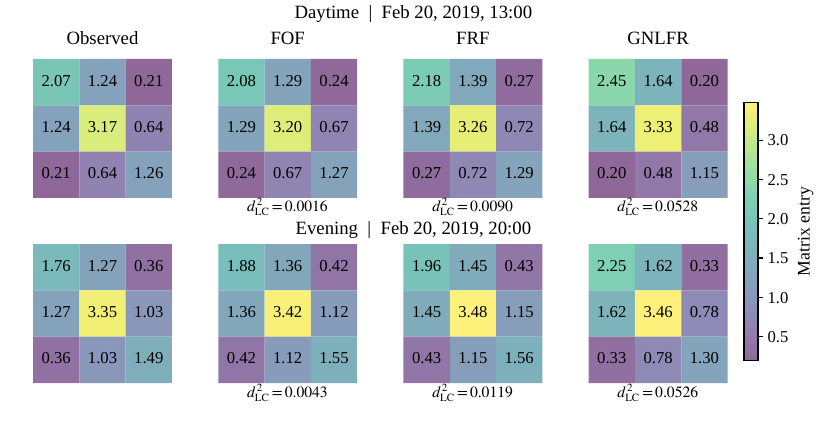}
\caption{Observed and predicted SPD matrices at 13:00 and 20:00 on
20 February 2019 for FOF, FRF and GNLFR.}
\label{fig:taxi-spd-comparison}
\end{figure}

Let $\mathcal T$ be the set of test hours, $\widehat S_t$ the predicted
SPD matrix, and $\bar S_{\mathcal T}$ the Fr\'echet mean of the observed
test responses. The MSE and Fr\'echet $R^2$ are
\[
 \operatorname{MSE}=\frac{1}{|\mathcal T|}
   \sum_{t\in\mathcal T}d_{LC}^2(S_t,\widehat S_t),
 \qquad
 R^2=1-\frac{\sum_{t\in\mathcal T}d_{LC}^2(S_t,\widehat S_t)}
                  {\sum_{t\in\mathcal T}d_{LC}^2(S_t,\bar S_{\mathcal T})}.
\]
FOF, FRF and GNLFR yield $(\mathrm{MSE}\times10^{2},R^2)$ values of
$(1.104,0.8692)$, $(1.239,0.8532)$ and $(6.436,0.2374)$, respectively.
FOF has the lowest test MSE, approximately $10.9\%$ below FRF.
Figure~\ref{fig:taxi-spd-comparison} illustrates two forecasts from the
test set.

\FloatBarrier

\AtBeginEnvironment{thebibliography}{\markboth{REFERENCES}{References}}
\bibliographystyle{plainnat}
\bibliography{paper-ref}

\appendix
\section*{Appendix 1}
\markboth{APPENDIX 1}{Appendix 1}
\setcounter{equation}{0}
\renewcommand{\theequation}{A\arabic{equation}}
\renewcommand{\thelemma}{A\arabic{lemma}}
\renewcommand{\theproposition}{A\arabic{proposition}}

\subsection*{Ridge approximation and Wasserstein bracketing}

\begin{lemma}[Ridge approximation]
\label{lem:ridge-density}
Suppose that $S$ is compact and that there are a real Hilbert space
$\mathbb H$ and a map $\Phi_\mu:\mathcal M_X\to\mathbb H$ such that
$\Phi_\mu(\mu)=0$, $\Phi_\mu$ is continuous and injective on $S$,
$\Phi_\mu(\mathcal M_X)$ is convex, and
\begin{equation*}
 h_\mu(x,\beta)
 =\langle\Phi_\mu(x),\Phi_\mu(\beta)\rangle_{\mathbb H},
 \qquad x\in S,\quad\beta\in\mathcal M_X.
\end{equation*}
Then
\begin{equation*}
 \overline{\mathcal G_\mu}^{\,L^2(P_X)}=L^2(P_X).
\end{equation*}
For compact $S$, the same conclusion holds if $\mathcal M_X$ is a connected complete
$p$-dimensional Riemannian manifold without boundary and the score is defined
using a Borel minimizing-log selection
$Z_\mu=\operatorname{Log}^{\mathrm{sel}}_\mu:\mathcal M_X\to
T_\mu\mathcal M_X$ satisfying
$Z_\mu(\mu)=0$, $\operatorname{Exp}_\mu\{Z_\mu(x)\}=x$, and
$\|Z_\mu(x)\|_\mu=d_X(\mu,x)$.

The conclusion applies to the following spaces:
\begin{enumerate}
\item $\mathcal M_X=\mathcal W_2(\mathbb R)$ with its $W_2$ metric;
\item $\mathcal M_X=\mathcal S_{++}^d$ with the Log--Cholesky metric;
\item any connected complete finite-dimensional Riemannian manifold equipped
with the Borel minimizing-log selection used to define $h_\mu$, including
every finite-dimensional Hadamard manifold;
\item $\mathcal M_X=\mathbb S^d$ with its geodesic metric, provided that a
minimizing geodesic from $\mu$ to $-\mu$ is chosen measurably in the score.
\end{enumerate}
For spherical predictors, no selection rule is needed if
$\mathcal M_X=\overline B_{\mathbb S^d}(\mu,\rho)$ with
$0<\rho<\pi/2$. Here $\mathbb S^d$ is the unit sphere in
$\mathbb R^{d+1}$ and $\overline B_{\mathbb S^d}(\mu,\rho)$ denotes its
closed geodesic ball of centre $\mu$ and radius $\rho$.
\end{lemma}

\begin{proposition}[Fixed-anchor Wasserstein bracketing]
\label{prop:wass-hpfd}
Let $\mathcal M_X=\mathcal W_2(\mathbb R)$, fix a deterministic anchor
$\mu\in\mathcal M_X$, and retain every direction $\beta\in\mathcal M_X$.
Write $Q_x$ for the quantile function of $x$. Suppose that there are
deterministic nested linear subspaces $V_m\subset L^2(0,1)$ and constants
$C_0\ge1$ and $r>0$ such that
\begin{equation}
 V_m\subseteq V_{m+1},\qquad
 \dim(V_m)\le m,\qquad
 \Pr(Q_X\notin V_m)\le C_0m^{-r},\qquad m\ge1.
 \label{eq:wass-hpfd}
\end{equation}
Then there is a constant $C<\infty$ such that
\begin{equation*}
 \log N_{[]}\{\varepsilon,\mathcal H_\mu,L^1(P_X)\}
 \le C\varepsilon^{-1/r}\log(C/\varepsilon),
 \qquad 0<\varepsilon<1.
\end{equation*}
Consequently, for every fixed $q>1/r$ there is a constant $C_H<\infty$ such
that \eqref{eq:A2-bracketing} holds.
\end{proposition}

\clearpage
\thispagestyle{plain}
\section*{Appendix 2}
\markboth{APPENDIX 2}{Appendix 2}
\subsection*{Fr\'echet oblique forest algorithm}

Algorithm~\ref{alg:frechet_oblique_forest} combines the nodewise rule in
Algorithm~\ref{alg:optimal_split_node} with the weights in
\eqref{eq:forest-weights-main}. The theory uses fixed-depth trees grown on
subsamples without replacement, with nonempty children at every split.
For the simulations, we replace subsampling with bootstrap sampling and
allow the stopping rules below.

\begingroup
\singlespacing
\RestyleAlgo{ruled}
\begin{algorithm}[H]
\caption{Fr\'echet Oblique Forest}
\label{alg:frechet_oblique_forest}
\small
\DontPrintSemicolon
\LinesNotNumbered
\textbf{Input:} Data $\mathcal D_n=\{(X_i,Y_i)\}_{i=1}^n$, anchor $\mu$,
query point $x$, number of trees $B_n$, subsample size $N_n$, maximum depth
$K_n$, minimum split and leaf sizes $n_{\mathrm{split}}$ and
$n_{\mathrm{leaf}}$, minimum gain $\delta_{\mathrm{gain}}\ge0$, and a
rule for generating candidate directions $\mathcal A_t$.\;

\For{$b=1,\ldots,B_n$}{
Draw $\mathcal I^{(b)}\subseteq\{1,\ldots,n\}$ uniformly without replacement,
with $|\mathcal I^{(b)}|=N_n$.\;
Grow $T^{(b)}$ recursively. At a node $t$ with observation index set
$\mathcal I_t$, stop if its depth reaches $K_n$ or
$|\mathcal I_t|<n_{\mathrm{split}}$. Otherwise, generate $\mathcal A_t$
and apply Algorithm~\ref{alg:optimal_split_node} with minimum child size
$n_{\mathrm{leaf}}$. Make $t$ terminal if the algorithm returns no split
with gain exceeding $\delta_{\mathrm{gain}}$. For an accepted split
$(\widehat\beta_t,\widehat c_t)$, send $z$ left if
$h_\mu(z,\widehat\beta_t)\le\widehat c_t$ and right otherwise. Continue
in both children.\;
Let $\ell_b(x)$ be the terminal leaf containing $x$.\;
}

Average the within-leaf weights:
\[
\bar\alpha_i(x)=\frac1{B_n}\sum_{b=1}^{B_n}
 \frac{\mathbf 1\{i\in\mathcal I^{(b)},\ X_i\in\ell_b(x)\}}
      {n_{\ell_b(x)}^{(b)}}.
\]
Return the weighted Fr\'echet prediction $\widehat f_n(x)$ from
\eqref{eq:forest-weighted-objective}.\;
\end{algorithm}
\endgroup

\section*{Appendix 3}
\markboth{APPENDIX 3}{Appendix 3}
\subsection*{Additional simulation results}
\setcounter{table}{0}
\renewcommand{\thetable}{S\arabic{table}}

As in the main experiment, we use $n=200,500,800$ training observations,
an independent test sample of size $n'=n/4$, and 100 replications per setting. We
compare FOF, FRF and GNLFR in Examples~\ref{ex:distribution-distribution}
and~\ref{ex:two-spd-predictors}, adding geodesic optimal transport regression
(GOT; \citealp{zhu2026geodesic}) where applicable. In these
two examples, the tuning parameters for FOF, FRF and GNLFR are the same as
in Example~\ref{ex:spd-spd}; both forests use 100 trees and the same
bootstrap samples within each replication. Example~\ref{ex:spherical-functional}
compares FOF with methods tailored to Riemannian functional predictors.
For the independent test pairs $(X_i^{\mathrm{test}},Y_i^{\mathrm{test}})$,
prediction accuracy is evaluated by
\[
 \operatorname{MSE}=\frac{1}{n'}\sum_{i=1}^{n'}
 d_Y^2(Y_i^{\mathrm{test}},\widehat Y_i^{\mathrm{test}}),
\]
where $\widehat Y_i^{\mathrm{test}}$ is the prediction at $X_i^{\mathrm{test}}$.

We consider two distribution-to-distribution regression settings.
\setcounter{example}{0}
\renewcommand{\theexample}{S\arabic{example}}
\begin{example}[Probability-distribution predictors and responses]
\label{ex:distribution-distribution}
Take $\mathcal M_X=\mathcal M_Y=\mathcal W_2(\mathbb R)$ and equip both
spaces with the 2-Wasserstein metric. In both settings, distributions are
represented by $m=21$ quantiles on an equally spaced probability grid
from $0.001$ to $0.999$.

\textup{(i)} Set $\mu=\operatorname{Beta}(6,2)$ and generate
$\xi_{ij}\overset{\mathrm{iid}}{\sim}N(0,0.025^2)$,
$j=1,\ldots,10$. Construct the predictor quantiles from
\[
\widetilde Q_{X_i}(p)
 =Q_\mu(p)+\sum_{j=1}^{10}\xi_{ij}\sqrt{2}\sin(2\pi jp),
\]
and apply a cumulative maximum to ensure monotonicity. Let
$\beta_1=\operatorname{Gamma}(2,\mathrm{scale}=3)$,
$\beta_2=N(0,1)$, $\beta_3=\operatorname{Beta}(2,5)$ and
$h_k(X_i)=h_\mu(X_i,\beta_k)$. Generate
\[
\begin{split}
 m(X_i)&=\sin\{10h_1(X_i)\}+5\{10h_2(X_i)\}^2
        -5\tanh\{100h_3(X_i)\},\\
 Y_i&=N\{m(X_i)+\epsilon_i,1\},\qquad
 \epsilon_i\overset{\mathrm{iid}}{\sim}N(0,1),
\end{split}
\]
with $\epsilon_i$ independent of $X_i$.

\textup{(ii)} We adapt Model I.4 of
\citet{bhattacharjee2025nonlinear}. Independently generate
$a_i\sim\operatorname{Gamma}(2,\mathrm{rate}=1)$ and
$b_i\sim\operatorname{Gamma}(2,\mathrm{rate}=3)$, and set
$X_i=\operatorname{Beta}(a_i,b_i)$. Let
$\mu_1=\operatorname{Beta}(2,1)$,
$\mu_2=\operatorname{Beta}(2,3)$ and
$d_{ik}=W_2(X_i,\mu_k)$ for $k=1,2$. Conditional on $X_i$, draw
independently
\[
 \mu_{Y_i}\sim N\{\exp(4d_{i1}^2),0.1\},\qquad
 \sigma_{Y_i}\sim\operatorname{Gamma}
 (\mathrm{shape}=d_{i2}^2,\mathrm{scale}=d_{i2}),
\]
and set $Y_i=N(\mu_{Y_i},\sigma_{Y_i}^2)$. Thus, both the location and
scale of the response distribution depend nonlinearly on the predictor.
\end{example}

\begin{table}[!htb]
\centering
\caption{Mean squared prediction errors in Example~\ref{ex:distribution-distribution}(i),
comparing FOF with FRF, GNLFR and GOT. Monte Carlo standard
deviations appear in parentheses; the lowest MSE for each sample size is in bold}
\label{tab:distribution_distribution_mse}
\small
\begin{tabular*}{\textwidth}
{@{\extracolsep{\fill}}lcccc@{}}
\toprule
$n$ & FOF & FRF & GNLFR & GOT \\
\midrule
200 & \textbf{1.37} (0.33) & 1.46 (0.38) & 1.58 (0.42) & 2.35 (0.58) \\
500 & \textbf{1.26} (0.16) & 1.35 (0.18) & 1.53 (0.25) & 2.28 (0.34) \\
800 & \textbf{1.21} (0.13) & 1.31 (0.14) & 1.49 (0.20) & 2.30 (0.27) \\
\bottomrule
\end{tabular*}
\end{table}

In Table~\ref{tab:distribution_distribution_mse}, FOF gives the lowest
error at every sample size, with reductions of about $6$--$8\%$ relative
to FRF. At $n=800$, the reductions relative to GNLFR and GOT are
approximately $19\%$ and $47\%$, respectively.

\begin{table}[!htb]
\centering
\caption{Mean squared prediction errors in Example~\ref{ex:distribution-distribution}(ii),
comparing FOF with FRF, GNLFR and GOT. Monte Carlo standard
deviations appear in parentheses; the lowest MSE for each sample size is in bold}
\label{tab:distribution_distribution_i4_mse}
\small
\begin{tabular*}{\textwidth}
{@{\extracolsep{\fill}}lcccc@{}}
\toprule
$n$ & FOF & FRF & GNLFR & GOT \\
\midrule
200 & \textbf{0.189} (0.091) & 0.191 (0.097) & 0.256 (0.132) & 0.298 (0.207) \\
500 & 0.174 (0.038) & \textbf{0.172} (0.038) & 0.277 (0.089) & 0.302 (0.092) \\
800 & 0.169 (0.035) & \textbf{0.166} (0.035) & 0.269 (0.071) & 0.296 (0.074) \\
\bottomrule
\end{tabular*}
\end{table}

For setting (ii), Table~\ref{tab:distribution_distribution_i4_mse} shows
that FOF leads at $n=200$, while FRF is slightly better at $n=500,800$;
their MSEs differ by at most $0.004$. Both forests outperform GNLFR and
GOT throughout, with errors decreasing as the sample grows. At $n=800$, FOF
reduces MSE by approximately $37.0\%$ relative to GNLFR and $42.7\%$
relative to GOT.

We next consider regression with multiple predictors.
\begin{example}[Multiple predictors and object responses]
\label{ex:two-spd-predictors}
\textup{(i)} We adapt Scenario~2 of \citet{zhu2026geodesic}. Let
$\mathcal M_X=(\mathcal S_{++}^3)^2$ and
$\mathcal M_Y=\mathcal S_{++}^3$, with the Log--Cholesky metric on each
factor. Independently generate
$U_{i,1}\sim N_3\{(1,2,3)^\top,\Sigma_1\}$,
$U_{i,2}\sim N_3\{(-1,1,-1)^\top,\Sigma_2\}$ and
$\epsilon_i\sim N_3(0,0.5I_3)$, where
\[
\Sigma_1=\begin{pmatrix}1&-0.3&0.1\\-0.3&1&0.3\\0.1&0.3&1\end{pmatrix},
\qquad
\Sigma_2=\begin{pmatrix}1&0.6&0.5\\0.6&1&0.7\\0.5&0.7&1\end{pmatrix}.
\]
For $j=1,2$, put $X_{i,j}=U_{i,j}U_{i,j}^\top+I_3$. With
$v_i=0.7U_{i,1}+0.2U_{i,2}$, the response is
$Y_i=v_iv_i^\top+I_3+\epsilon_i\epsilon_i^\top$.
Each method receives only the observed predictors $(X_{i,1},X_{i,2})$.

\textup{(ii)} Let
$\mathcal M_X=\mathbb R\times\mathcal W_2(\mathbb R)
\times\mathcal S_{++}^3\times\mathbb S^2$ and
$\mathcal M_Y=\mathbb S^2$, where $\mathbb S^2$ is the unit sphere in
$\mathbb R^3$. Equip the four predictor factors with the Euclidean,
2-Wasserstein, Log--Cholesky and spherical geodesic distances,
respectively, and use the chordal response distance
$d_Y(y,y')=\|y-y'\|_2$. Independently generate
\[
 R_i,M_i,T_i,A_{i1},\ldots,A_{i6},V_{i1},V_{i2}
 \overset{\mathrm{iid}}{\sim}\operatorname{Unif}(-1,1).
\]
Set $X_i=(R_i,\nu_i,P_i,S_i)$, where
\[
 \nu_i=N\{M_i,(1+T_i/2)^2\},\qquad
 P_i=L_iL_i^\top,\qquad
 L_i=\begin{pmatrix}
 e^{A_{i1}/2}&0&0\\
 A_{i4}&e^{A_{i2}/2}&0\\
 A_{i5}&A_{i6}&e^{A_{i3}/2}
 \end{pmatrix}.
\]
Put $o=(0,0,1)^\top$ and
$S_i=\operatorname{Exp}_o\{(V_{i1},V_{i2},0)^\top\}$, where
\[
 \operatorname{Exp}_s(v)=\cos(\|v\|_2)s+
             \sin(\|v\|_2)v/\|v\|_2,
 \qquad v\in T_s\mathbb S^2,
\]
with its continuous extension at $v=0$. Define
\[
 a_i=\frac{A_{i4}+A_{i5}+V_{i1}+V_{i2}}{2},\qquad
 b_i=\frac12\sin(\pi R_i)+M_i^2+\frac{T_i^2}{4}-\frac5{12}.
\]
Here $M_i^2+T_i^2/4=W_2^2\{\nu_i,N(0,1)\}$, whose expectation is
$5/12$. Independently of the predictors and across $i$, draw
$Z_i\sim N_3(0,I_3)$ and set
\[
 m_i=\frac{(a_i,b_i,1)^\top}{\sqrt{1+a_i^2+b_i^2}},\qquad
 Y_i=\operatorname{Exp}_{m_i}\{0.2(I_3-m_im_i^\top)Z_i\}.
\]
The conditional Fr\'echet mean under the chordal metric is $m_i$.
For fitting, use the same 21-point quantile grid as in
Example~\ref{ex:distribution-distribution} and approximate squared
Wasserstein distances by mean squared quantile differences.
\end{example}

In setting \textup{(i)}, FOF-product treats the predictors as one object in the
unweighted product Log--Cholesky space. FOF-coordinate searches the two SPD predictors
separately and uses the split with the larger nodewise gain:
\[
\begin{aligned}
h_\times(x,\beta)&=\sum_{j=1}^2 h_{\mu_j}(x_j,\beta_j)
  &&\text{(FOF-product)},\\
(j_t,\widehat\beta_t,\widehat c_t)&\in
   \operatorname*{argmax}_{\substack{j\in\{1,2\}\\(\beta,c)\in\mathcal C_{t,j}}}
   \Delta_t^{(j)}(\beta,c)
  &&\text{(FOF-coordinate)}.
\end{aligned}
\]
Here $\mu_j$ is the training Fr\'echet mean of predictor $j$ and
$\mathcal C_{t,j}$ is the set of direction--threshold pairs evaluated
for that predictor at node $t$. The gain $\Delta_t^{(j)}(\beta,c)$
corresponds to the split $h_{\mu_j}(x_j,\beta)\le c$.
FRF, GNLFR and GOT use the same predictor pair and response samples.

\begin{table}[!htb]
\centering
\caption{Mean squared prediction errors in Example~\ref{ex:two-spd-predictors}(i),
comparing the two FOF variants with FRF, GNLFR and GOT. Monte Carlo standard
deviations appear in parentheses; the lowest MSE for each sample size is in bold}
\label{tab:two_spd_predictors_mse}
\small
\begin{tabular*}{\textwidth}
{@{\extracolsep{\fill}}lccccc@{}}
\toprule
$n$ & FOF-product & FOF-coordinate & FRF & GNLFR & GOT \\
\midrule
200 & \textbf{0.636} (0.102) & 0.673 (0.110) & 0.662 (0.106) & 0.977 (0.183) & 0.672 (0.108) \\
500 & \textbf{0.539} (0.051) & 0.565 (0.054) & 0.552 (0.054) & 0.977 (0.119) & 0.657 (0.061) \\
800 & \textbf{0.512} (0.033) & 0.532 (0.036) & 0.522 (0.033) & 1.001 (0.089) & 0.665 (0.044) \\
\bottomrule
\end{tabular*}
\end{table}

FOF-product performs best at every sample size in
Table~\ref{tab:two_spd_predictors_mse}, reducing MSE by $3.8$--$5.5\%$
relative to FOF-coordinate. Its improvement over FRF narrows from $3.9\%$
at $n=200$ to $1.9\%$ at $n=800$, while the reduction relative to GOT
grows from $5.3\%$ to $23.0\%$. Errors are $34.9$--$48.9\%$ lower than
those of GNLFR. Both FOF variants improve as the sample grows, whereas
GNLFR and GOT change little.

In setting \textup{(ii)}, FOF-product uses a product metric with weights
$\widehat s_j^{-2}$ and the corresponding score
\[
 \widehat s_j=\operatorname{median}_{1\le i<k\le n}
 d_j(X_{i,j},X_{k,j}),\qquad
 h_\times(x,\beta)=\sum_{j=1}^4\widehat s_j^{-2}
 h_{\mu_j}(x_j,\beta_j),
\]
where $d_j$ is the distance on modality $j$. The scales and Fr\'echet
anchors $\mu_j$ are computed once from the full training sample, and
each product direction is a complete training observation.
FOF-coordinate searches the four modalities separately using the same
nodewise gain criterion as in \textup{(i)}.

\begin{table}[!htb]
\centering
\caption{Mean squared prediction errors in Example~\ref{ex:two-spd-predictors}(ii),
comparing the two FOF variants with FRF and GNLFR. Monte Carlo standard
deviations appear in parentheses; the lowest MSE for each sample size is in bold}
\label{tab:multimodal_sphere_mse}
\small
\begin{tabular*}{\textwidth}
{@{\extracolsep{\fill}}lcccc@{}}
\toprule
$n$ & FOF-product & FOF-coordinate & FRF & GNLFR \\
\midrule
200 & 0.197 (0.026) & \textbf{0.165} (0.021) & 0.216 (0.028) & 0.244 (0.037) \\
500 & 0.160 (0.013) & \textbf{0.137} (0.012) & 0.184 (0.015) & 0.251 (0.025) \\
800 & 0.143 (0.011) & \textbf{0.124} (0.009) & 0.167 (0.012) & 0.239 (0.018) \\
\bottomrule
\end{tabular*}
\end{table}

For the mixed predictors in Table~\ref{tab:multimodal_sphere_mse},
FOF-coordinate is the most accurate method, with MSE reductions of about
$13$--$16\%$ relative to FOF-product and $23$--$26\%$ relative
to FRF. At $n=800$, its reduction relative to GNLFR is $48.1\%$.
Both FOF variants and FRF improve as the training sample grows.

Finally, we consider non-Euclidean functional predictors with scalar responses.
\begin{example}[Spherical functional predictors and scalar responses]
\label{ex:spherical-functional}
We adapt the spherical setting of Model~I in \citet{chen2026intrinsic}.
Each predictor is a curve $X_i\colon[0,1]\to\mathbb S^2$, observed at
$t_j=(j-1/2)/m$, $j=1,\ldots,m$, with $m=100$, and
$\mathcal M_Y=\mathbb R$ with the Euclidean distance. Fix the reference curve
\[
 \mu(t)=\{\sin\varphi(t)\cos\theta(t),\,
 \sin\varphi(t)\sin\theta(t),\,\cos\varphi(t)\}^{\top},
\]
where $\theta(t)=2t^2+4t+1/2$ and
$\varphi(t)=(t^3+3t^2+t+1)/2$. Let $E(t)$ be the spherical orthonormal
frame used in the Supplementary Material of \citet{chen2026intrinsic},
and let $A$ be a random $2\times2$ orthogonal matrix fixed throughout.
For $\ell=1,\ldots,10$, set
\[
 \phi_{2\ell-1}(t)=f_\ell(t)E(t)Ae_1,\qquad
 \phi_{2\ell}(t)=f_\ell(t)E(t)Ae_2,
\]
where $e_1,e_2$ are the standard coordinate vectors in $\mathbb R^2$ and
$f_{2r-1}(t)=\sqrt{2}\sin(2\pi rt)$,
$f_{2r}(t)=\sqrt{2}\cos(2\pi rt)$, $r=1,\ldots,5$.
Independently generate
\[
 V_i(t)=\frac12\sum_{k=1}^{20}\sqrt{2k^{-1.2}}\,U_{ik}\phi_k(t),
 \qquad X_i(t)=\operatorname{Exp}_{\mu(t)}\{V_i(t)\},
\]
where $U_{ik}\overset{\mathrm{iid}}{\sim}\operatorname{Unif}(-\pi/4,\pi/4)$.
The amplitude ensures $\sup_t\|V_i(t)\|_2<\pi$, so that
$V_i(t)=\operatorname{Log}_{\mu(t)}X_i(t)$.
With $B(t)=\sum_{k=1}^{20}(1.5/k^2)\phi_k(t)$, generate
\[
 S_i=\frac1m\sum_{j=1}^{m}\langle V_i(t_j),B(t_j)\rangle,
 \qquad Y_i=S_i+\exp(S_i)+\epsilon_i,
\]
where $\epsilon_i\overset{\mathrm{iid}}{\sim}N(0,0.4^2)$ is independent of
the predictors. The prediction methods receive only the observed curves.
\end{example}

We compare FOF-product with the truncated iRF-WIRE and iRF-WDR methods of
\citet{chen2026intrinsic}, each followed by ordinary linear regression
with an intercept (LR) or Gaussian local linear regression (LL).
Both dimension-reduction methods use truncation level $K=20$ and
structural dimension $d=1$. For LL, we standardize the estimated score
using the training data and set $h=c\,n_{\mathrm{fit}}^{-1/5}$, where
$n_{\mathrm{fit}}$ is the fitting sample size. We select $c$ from 15
logarithmically spaced values in $[0.25,4]$ by five-fold cross-validation,
refitting the mean curve, functional principal components, supervised
direction and standardization within each training fold.

\begin{table}[!htb]
\centering
\caption{Mean squared prediction errors in Example~\ref{ex:spherical-functional},
comparing FOF-product with truncated iRF-WIRE and iRF-WDR followed by
local linear (LL) or linear regression (LR). Monte Carlo standard
deviations appear in parentheses; the lowest MSE for each sample size is in bold}
\label{tab:spherical_functional_mse}
\small
\begin{tabular*}{\textwidth}
{@{\extracolsep{\fill}}lccccc@{}}
\toprule
& FOF-product & \multicolumn{2}{c}{iRF-WIRE} & \multicolumn{2}{c}{iRF-WDR} \\
\cmidrule(lr){3-4}\cmidrule(lr){5-6}
$n$ & & LL & LR & LL & LR \\
\midrule
200 & 0.1885 (0.0405) & \textbf{0.1873} (0.0383) & 0.1967 (0.0408) & 0.1913 (0.0404) & 0.1999 (0.0413) \\
500 & 0.1764 (0.0220) & \textbf{0.1720} (0.0220) & 0.1844 (0.0249) & 0.1726 (0.0225) & 0.1852 (0.0252) \\
800 & 0.1735 (0.0167) & \textbf{0.1674} (0.0176) & 0.1800 (0.0174) & 0.1676 (0.0175) & 0.1800 (0.0173) \\
\bottomrule
\end{tabular*}
\end{table}

In Table~\ref{tab:spherical_functional_mse}, iRF-WIRE-LL yields the lowest
MSE at every sample size, with iRF-WDR-LL close behind at $n=500,800$.
FOF-product's errors are approximately $0.7\%$, $2.6\%$ and $3.7\%$ higher
at $n=200,500,800$, respectively, and remain below those of both LR-based
competitors. The same product-space splitting rule thus gives competitive
accuracy for spherical curves without a preliminary supervised
dimension-reduction step.

\FloatBarrier

\end{document}